\documentclass[12pt,preprint]{aastex}
\def\vs{$v \sin i$}
\def\kms{${\rm km s}^{-1}$}

\shorttitle{$V \sin i$ of OB stars from He\,{\sc i} lines}
\shortauthors{Daflon et al.}

\begin{document}

\title{The Projected Rotational Velocity Distribution of a Sample of OB stars from 
a Calibration based on Synthetic He\,{\sc i} lines}
                                                                                         
\author{Simone Daflon}
\affil{Observat\'orio Nacional/MCT, R, Gal Jos\'e Cristino 77,
20921-400 Rio de Janeiro/RJ, Brazil; daflon@on.br}
\author{Katia Cunha
\footnote{on leave from Observat\'orio Nacional/MCT, Rio de Janeiro, Brazil}}
\affil{National Optical Astronomy Observatory, P.O. Box 26732, Tucson, AZ
85726, USA; kcunha@noao.edu}
\author{Francisco X. de Ara\'ujo}
\affil{Observat\'orio Nacional/MCT, R, Gal Jos\'e Cristino 77,
20921-400 Rio de Janeiro/RJ, Brazil; araujo@on.br}
\author{Sidney Wolff}
\affil{National Optical Astronomy Observatory, P.O. Box 26732, Tucson,
AZ 85726, USA; swolff@noao.edu}
\author{Norbert Przybilla}
\affil{Dr Reimes Sternwarte, Bamberg, Germany; przybilla@sternwarte.uni-erlangen.de}
                                                                                         
\begin{abstract}
We derive projected rotational velocities (\vs) for a sample of 156 Galactic OB star members
of 35 clusters, H\,{\sc ii} regions, and associations. 
The He\,{\sc i} lines at $\lambda\lambda$4026, 4388, and 4471\AA\ were analyzed 
in order to define a calibration of the synthetic He\,{\sc i} full-widths at half 
maximum versus stellar \vs.  A grid of synthetic spectra of He\,{\sc i} line profiles 
was calculated in non-LTE using an extensive helium model atom and updated atomic data. 
The \vs's for all stars were derived using the He\,{\sc i} FWHM calibrations but also, 
for those target stars with relatively sharp lines, \vs\ values were obtained from best fit synthetic
spectra of up to 40 lines of C\,{\sc ii}, N\,{\sc ii}, O\,{\sc ii}, Al\,{\sc iii}, 
Mg\,{\sc ii}, Si\,{\sc iii}, and S\,{\sc iii}.  This calibration is a useful and efficient tool for 
estimating the projected rotational velocities of O9-B5 main-sequence stars.  The distribution of 
\vs\ for an unbiased sample of early B stars in the unbound association Cep OB2 is consistent 
with the distribution reported elsewhere for other unbound associations.
\end{abstract}


\section{Introduction}

The distribution of projected rotational velocities (\vs) of stars 
as a function of spectral type shows that the B-type stars have the largest 
average \vs\ values among all main sequence stars. This observational result makes 
rotation especially important for early-type stars because it may affect the 
star's evolution and important stellar characteristics, such as
the surface abundances. Moreover, there is growing evidence that
the \vs\ distribution of OB stars is somehow related to the physical 
characteristics of the region in which they formed. These two 
important issues were addressed recently by, for example, 
\citet{heg06a,heg06b} and \citet{wol07}.

An approach that is frequently employed in order to obtain estimates of the projected
rotational velocity consists of measuring the full-widths at half maximum
(FWHM) of He\,{\sc i} lines and adopting relationships  
between FWHM and \vs\ established from observational data such as given by 
\citet{sle75,sle82} and \citet{how97}. Alternatively,
the projected rotational speed can be obtained from the comparison between 
observational and synthetic profiles of He and metallic lines.

The goals of this contribution are to derive \vs's for a large sample of main 
sequence OB stars - most of them given for the first time in the literature - 
and to present a calibration for \vs\ versus the synthetic  full-widths at half 
maximum  of He\,{\sc i} lines at  $\lambda\lambda$4026, 4388, and 4471 \AA. 
In Section 2 we briefly describe 
our observational runs and data reduction procedures. Section 3 gives 
the \vs\ determination from the spectral synthesis of metal lines, a 
method that we only employed for stars with sharp lines. Section 4 details 
and discusses our new FWHM versus \vs\ calibration, while in Section 5 we 
present our \vs\ determinations. Finally, in Section 6 we discuss our
results, paying special attention to the case of Cep OB2 association and
outline some conclusions.

\section{Observations}\label{obs}

The observational data for this study are high resolution spectra of late O and early B
type stars that were obtained during several observing runs between the years 1994 and 2000.
The sample comprises 156 targets belonging to 35 open
clusters, H\,{\sc ii} regions, and OB associations.  Table~\ref{main} lists
 the observed stars and their cluster membership (Columns 1 and 2).
The sample of OB stars studied here for rotation was observed with the main goal  of 
analysing radial metallicity gradients of young stars in the Galactic disk \citep{pap7}.

The northern stars were observed with the McDonald Observatory telescopes
(University of Texas, Austin). We obtained echelle spectra (R=60,000 and
spectral coverage from $\sim$4200 to $\sim$4600\AA)
with the 2.1-m telescope and the Sandiford Cassegrain echelle spectrometer. Lower
resolution spectra (R=12,000 centered at 4340~\AA) were obtained 
with the 2.7-m telescope and a coud\'e spectrometer. Details of
the observations and data reduction can be found in papers by
Daflon, Cunha \& Becker (1999) and by Daflon et al. (2001a,b; 2003). 

\begin{figure}
\epsscale{.80}
\plotone{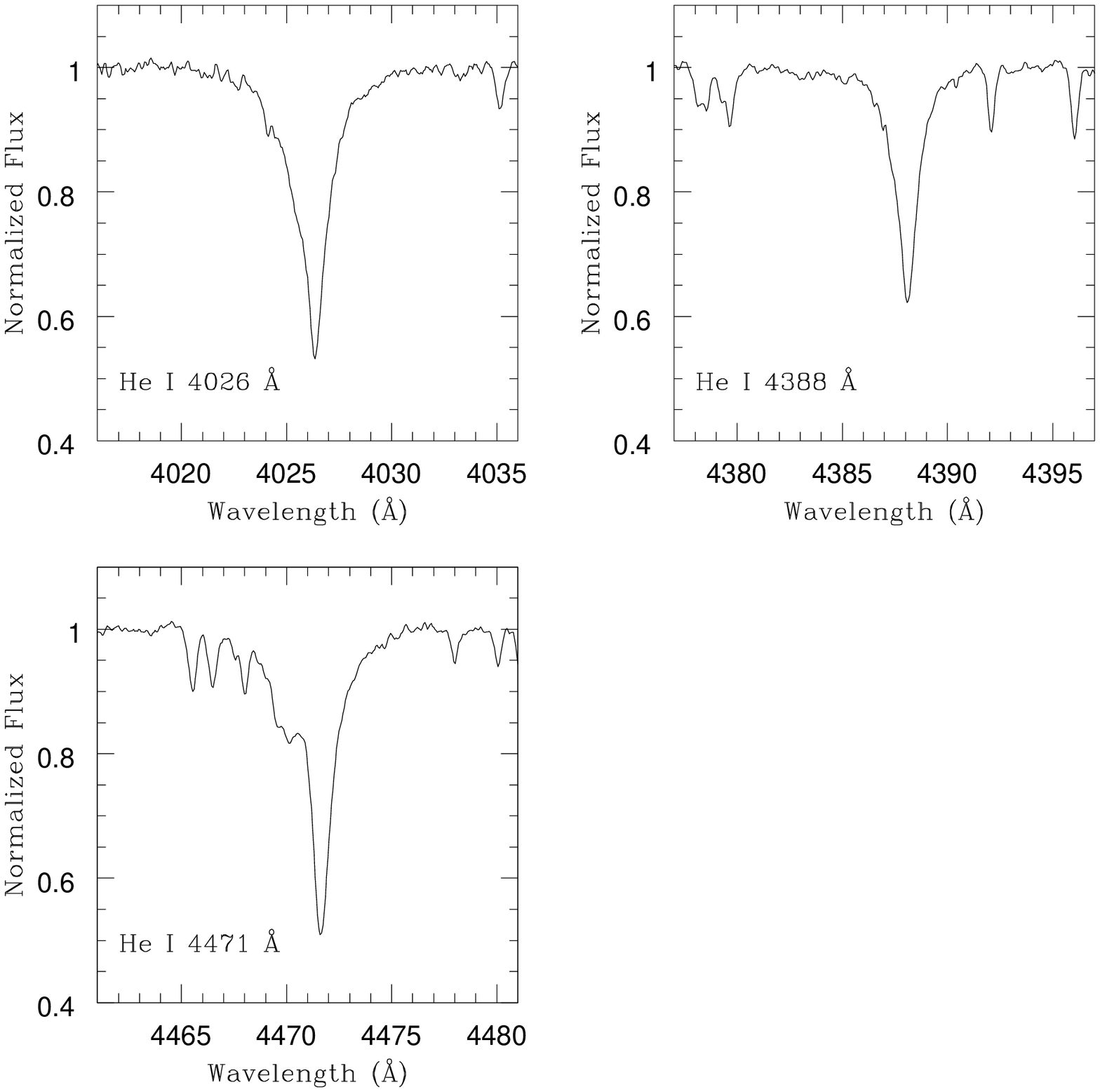}
\caption{Observed profiles of He\,{\sc i} lines at  $\lambda\lambda$4026, 4388 and 
4471~\AA \ for star Sh2-47 3. \label{obsf}}
\end{figure}

High resolution spectra for the southern stars in our sample were obtained
with the European Southern Observatory 1.52-m telescope at La Silla, Chile, coupled with FEROS
(Fiber fed Extended Range Optical Spectrograph), covering 3900-9200\AA \ and
with a resolving power $\lambda / \Delta\lambda$=48,000.
(See Daflon, Cunha \& Butler 2004a,b  for more details).
In Figure~\ref{obsf} we show sample FEROS spectra for the target star
Sh2-47 3 in three spectral regions corresponding to the He\,{\sc i} lines at 
$\lambda\lambda$4026, 4388, and 4471\AA.

\section{$V \sin i$ Determination via Spectrum Synthesis of Metal Lines}\label{metals}

Stars with sharp spectral lines are best suited for abundance analyses via either
equivalent width measurements or spectrum synthesis. When line spectrum synthesis is
adopted to derive the elemental abundances, the corresponding value of the stellar 
projected velocity is also obtained. 
A significant fraction of the targets studied here (69 stars) had low enough values 
of rotational projected velocities so that their abundances could be derived
via spectrum synthesis of a number of spectral lines. 
The synthetic spectra were computed in non-LTE using the codes {\sc detail} \citep{gid81} 
and {\sc surface} \citep{beg85} plus line-blanketed LTE {\sc atlas9} model atmospheres \citep{kur93}.
The model atoms adopted in the calculations were taken from Eber \& Butler (1988 -- C\,{\sc ii}), 
Becker \& Butler (1989 -- N\,{\sc ii}), Becker \& Butler (1988 -- O\,{\sc ii }),
Dufton {\it et al.} (1986 -- Al\,{\sc iii}), Przybilla {\it et al.} (2001 -- Mg\,{\sc ii}),
Becker \& Butler (1990 -- Si\,{\sc iii}), and Vrancken, Butler \& Becker (1996 -- S\,{\sc iii}).
Stellar rotation is an important broadening mechanism that was included in the
computations of the synthetic spectra for the sample stars.
The computations also included limb darkening using a linear limb
darkening coefficient \citep{wer85}, a Gaussian broadening function
corresponding to the instrumental profile and the
macroturbulent velocity. We note that in the calculations of the
synthetic profiles of the target stars, which are on the main
sequence, we set the macroturbulence value to zero.
(Assuming a non-zero macroturbulence may be important   when one is interested in fitting 
wings of very sharp line profiles.)

\begin{figure}
\epsscale{.80}
\plotone{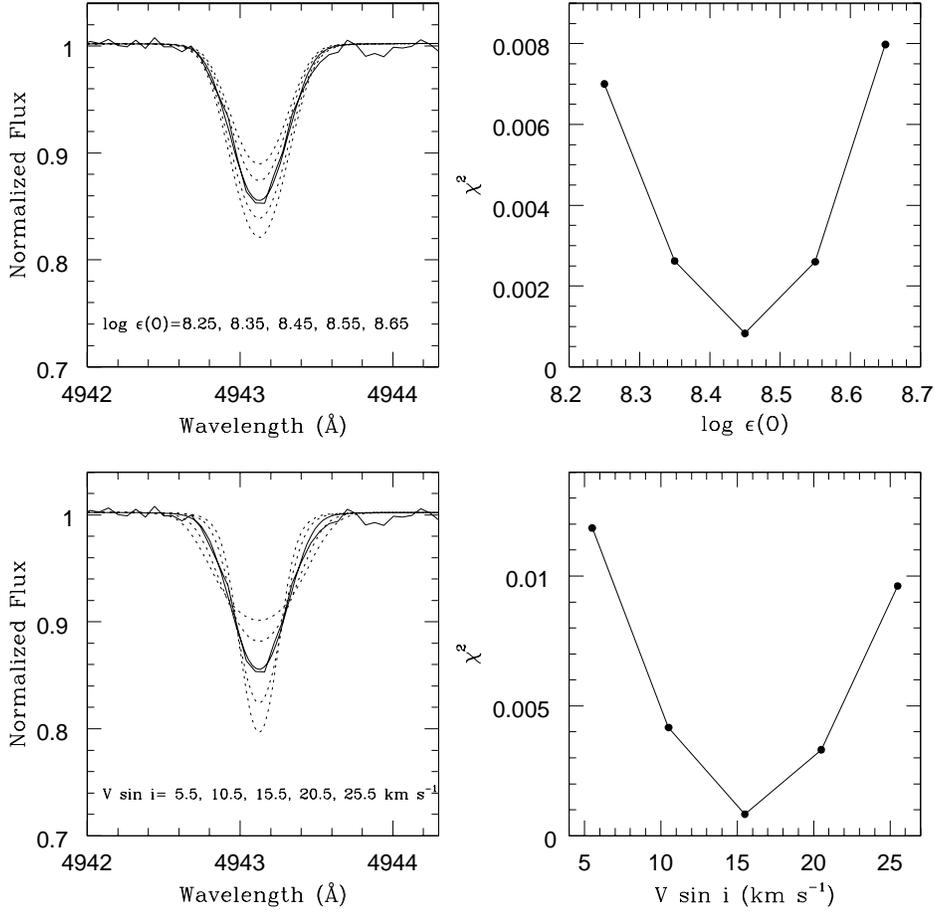}
\caption{Comparison between observed and non-LTE synthetic
profiles for one of the spectral regions containing the O\,{\sc ii} line at
4943~\AA. The observed
spectrum is for the target star Sh 2-47 3 and the synthetic profiles were
calculated for different sets of parameters.
{\it Top}: on the left panel we show synthetic profiles calculated for
five values of oxygen abundances (indicated in the figure).
The best fit oxygen abundance is derived for $\log \epsilon (O)$=8.45
(represented by the solid line).  In the right panel we present the variation of $\chi^2$ as
a function of oxygen abundance.  {\it Bottom}: same for different values of \vs\ 
varying from 5.5 to 25.5 \kms. \label{oii}}
\end{figure}

The stellar abundances and \vs's were derived from comparisons between observed and model
spectra. The best fit to the observed profile of each studied line was obtained from 
a $\chi^2$-minimization, which allowed for variation of the elemental abundance and \vs\
value. In order to illustrate the procedure adopted in order to find the best
fit synthetic spectrum for one studied oxygen line in one sample star 
we show, in Figure~\ref{oii}, observed spectra and synthetic profiles of the O\,{\sc ii} 
line at $\lambda$4943\AA\ for Sh2-47 3. In the top left panel five syntheses are shown
for different oxygen abundances and for one value of \vs = 15.5 \kms\
(indicated from the $\chi^2$-minimization shown in the right bottom panel);
the best fit oxygen abundance is obtained (represented by the synthetic profile
with a solid line in the top left panel) and it was derived via $\chi^2$-minimization (top right panel).
The bottom left panel illustrates the sensitivity of the profiles to the variation of the
\vs\ values ranging, in this example, between 5.5 and 25.5~\kms.

For each target star \vs\ values were obtained from synthetic fits, as described above,
of up to 40 spectral lines of  C\,{\sc ii}, N\,{\sc ii}, O\,{\sc ii }, Al\,{\sc iii},
Mg\,{\sc ii},  Si\,{\sc iii}, and S\,{\sc iii}. A list of the sample lines
analyzed for each ion can be found in \citet{pap7} and references therein. 
The \vs\ corresponding to the weak metal lines ($<V \sin i_M>$; 
listed in column4 of Table~\ref{main})
for each target is the average of all values obtained from the individual
line syntheses, and the line-to-line scatter represents the \vs\ uncertainties.
The stellar abundance results have been published in Daflon, Cunha \& Becker (1999); Daflon et al. 
(2001a,b; 2003); and Daflon, Cunha \& Butler (2004a,b). 
The overall analysis of the abundance distribution in the
Galactic disk, and in particular a discussion of the metallicity gradients
is presented in Daflon \& Cunha (2004).

\section{$V \sin i$ Determination via Measurements of Full-Widths at Half 
Maxima of He\,{\sc i} lines - A Calibration of \vs\ Based on the Synthetic 
FWHM of He\,{\sc i} Lines}\label{Helium}

For most stars in our sample, the determination of \vs\ from
the spectrum synthesis of a large number of weak metal lines, as discussed above, was not 
performed because of their high rotation.  A better alternative to obtain \vs's for early-type 
stars with a large range of rotation is to rely on the strong He\,{\sc i} lines, which are 
present in their spectra.  In order to obtain the \vs\ for all stars in our sample in a 
homogeneous way, we computed a grid of synthetic spectra of He\,{\sc i} lines in non-LTE, 
which we describe below. 


We constructed a grid of non-LTE synthetic He\,{\sc i} profiles at $\lambda\lambda$4026, 4388, 
and 4471\AA\ based on model atmospheres with a range in effective temperatures and for a variety 
of projected rotational velocities.
The He model atom adopted in the computations of  synthetic spectra is fairly complete, and it is
discussed in detail in Przybilla (2005; refer to that paper for details of the atomic data):
it includes all states up to the principal quantum number n=5 for He\,{\sc i} and all the 
levels/transitions up to n=20 for  He\,{\sc ii}. 
Line broadening for the He lines was treated according to the theory of \citet{sha69} for the 
lines 4026 and 4388\AA\ and \citet{bcs69} for the line 4471 \AA.
This model atom has been tested for helium formation in OB stars by \citet{nep07}, in a similar
approach than ours and they found  excellent fit quality of the He
profiles over the parameter range of our sample.

\begin{figure}
\epsscale{.80}
\plotone{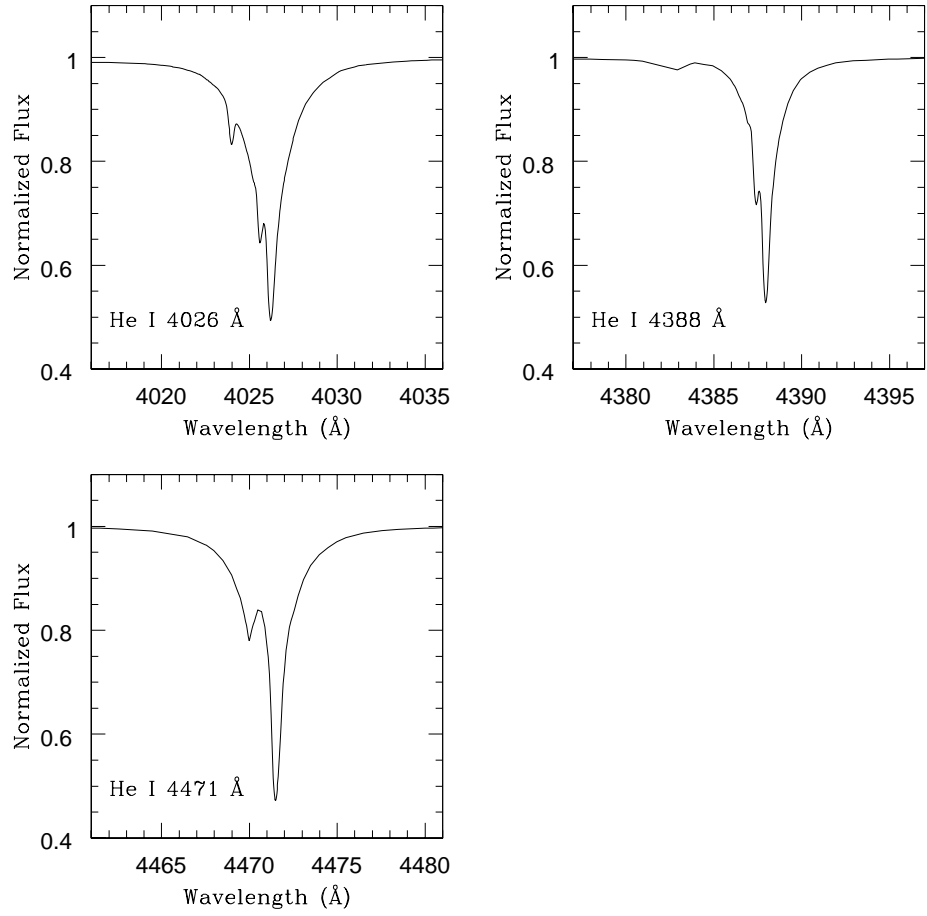}
\caption{Theoretical profiles non-convolved for rotation of He\,{\sc i} lines at  
$\lambda\lambda$4026, 4388 and 4471\AA, 
computed for $T_{\rm eff}=$25,000~K, $\log g$=4.0 and $\xi$=5.0~\kms. \label{prof}}
\end{figure}

The grid of He\,{\sc i} synthetic profiles was calculated using Kurucz model atmospheres 
\citep{kur93} with solar composition and for
effective temperatures bracketing spectral types roughly between B5 and O9, or
$T_{\rm eff}$=15,000, 20,000,  25,000, and 30,000~K. The surface gravities adopted for all the models
were $\log g = 4.0$ (corresponding to non-evolved stars) 
and the microturbulent velocities $\xi$ were kept constant at $\xi$=5~\kms, which is a typical value 
for the microturbulent velocity found in abundance studies of main-sequence OB stars (e.g. 
Dufton et al. 2006).
In Figure~\ref{prof} we show, as an example, unbroadened synthetic profiles for the selected 
He\,{\sc i} lines calculated using a model atmosphere with $T_{\rm eff}$=25,000~K.
The He\,{\sc i} synthetic profiles were then convolved for \vs\ values ranging from zero to 400~\kms, 
in steps of 50~\kms and with instrumental profiles corresponding to a resolution R=50,000. 

The FWHM of the convolved theoretical He\,{\sc i} line profiles were measured using the {\sc iraf}
package splot.
The continuum level was marked at the line center and the half-width of the red wing was measured
at the half maximum. The obtained value was doubled in order to derive the full-width at the half
maximum.  This procedure was adopted because the blue wings of the  He\,{\sc i} lines are
disturbed by secondary profiles such as $\lambda\lambda$ 4023.98 and 4026.2\AA \ of He\,{\sc i}
(and 4025.61\AA\ of He\,{\sc ii}, which may have some effect in the hottest models only)
and $\lambda\lambda$  4387.00 and 4469.96\AA \ of [He\,{\sc i}] (see Figure~\ref{prof}).
These weaker components are detached from the ``main'' He\,{\sc i} profiles for stars with low
\vs\ but become blended  as \vs\ increases. All theoretical profiles were measured consistently.
We note that for \vs$\leq$50~\kms, the FWHM measured by doubling the half-width is slightly larger than
the full-width at half maximum measured directly, especially for the line 4471\AA.
This difference tends to zero for \vs\ $>$ 100~\kms.

\begin{figure}
\epsscale{.80}
\plotone{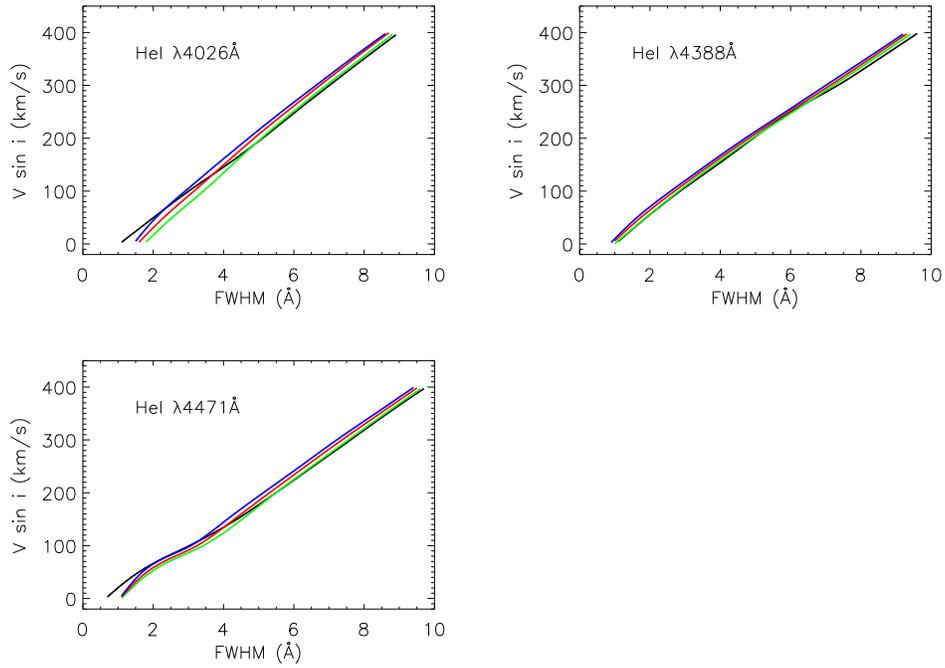}
\caption{Calibration of synthetic FWHM versus stellar projected roational velocity for sample  
He\,{\sc i} lines specified in the figure. The calculations
of synthetic  He\,{\sc i} profiles adopted model atmospheres with
 $T_{\rm eff}$= 15,000K (black curve);  $T_{\rm eff}$=20,000K (green curve)
 $T_{\rm eff}$=25,000K (red curve) and  $T_{\rm eff}$=30,000K (blue curve).
The calculations adopted an instrumental broadening
corresponding to R=50,000.   \label{calib}}
\end{figure}

The synthetic FWHM of the He\,{\sc i} line profiles for the 4 grid effective temperatures
are listed in Table~\ref{grid}. In Figure ~\ref{calib}
we show the variation of these theoretical FWHM as a function of \vs; 
all calculations presented in this figure adopted R=50,000 as the broadening of the instrumental 
profile because this value corresponds roughly to the resolution of our data. 
We note that whilst the curves for $T_{\rm eff}$=20,000, 25,000, and 30,000~K
(green, red and blue curves, respectively) run nearly paralell to each other, the
black curve for $T_{\rm eff}$=15,000~K presents a different behaviour as this curve
crosses the curves for the higher temperatures. This effect probably results from 
the fact that, for the  three  He\,{\sc i} lines studied here, the line intensity as
a function of spectral type reaches its maximum for B2V \citep{did82}, which corresponds 
to roughly an effective temperature of $\sim$20,000~K.

For the sake of completeness, we also calculated results for R=10,000.  As expected, 
diagrams of \vs\ versus FWHM calculated for an instrumental resolution of R=10,000
are very similar to the ones for R=50,000, with significant differences appearing only for 
\vs$\leq$50~\kms.  For a higher \vs, the rotational broadening becomes more important than 
the instrumental broadening and dominates the line profile. Thus, for moderate to high rotation, 
the \vs\ values derived from the medium- and high-resolution calibrations are virtually 
the same. For example, considering $T_{\rm eff}$=25,000~K and a measured FWHM 
of 2~\AA \ for the line 4026\AA, one would obtain \vs\ of 17 and 31~\kms\ for R=10,000 and 
50,000, respectively. For a measured FWHM$\sim$3~\AA, the derived \vs\  from both calibrations 
is $\sim$90~\kms.  

It should be noted that our calculations ignore the flattening of the stellar photosphere and 
gravitational 
darkening, which is a flux reduction toward the equator due to the decrease of the local 
 $T_{\rm eff}$ from the pole to the equator. These mechanisms were carefully re-analysed recently by 
\citet{toh04} and \citet{fre05}, among others. For instance, Table 2 of the \citet{toh04} paper
shows that for  $i=90^{\arcdeg}$ and $ve/vc=0.95$, where $ve$ is the equatorial rotational velocity
and $vc$ is the velocity at which centrifugal forces balance Newtonian gravity at the equator, \vs\ 
determinations that neglect these effects can be underestimated by
12 to 33\% (depending on the spectral type) when the He\,{\sc i} $\lambda$4471\AA \ line is 
used. These effects rapidly become negligible as the rotational rate
decreases, and only three of the stars in our sample have projected rotational
velocities that exceed 300~\kms\ or equivalently about three-quarters of the critical velocity in the 
mass range surveyed here.  Therefore, the effects of gravity darkening and flattening of 
the photosphere should be neglible for the vast majority of stars in our sample.  
While one should not employ our calibration for a sample of very rapidly rotating 
stars, such as Be-type stars, it can be safely used in a sample of OB stars like the one analyzed in
this study.

One interesting application of the gravity darkening effect
is the analysis of widhts of lines formed at different temperatures
and thus in different latitudes over the stellar disk.
As a result, the lines produced by ions with higher ionization
potentials tend to form closer to the poles \citep{fre04}.
In practice, the lines formed in different latitudes, implying in different
inclination angles $i$, present different widths, for the same \vs\ value.
An example of such analysis can be found in \citet{seb87},
who compared He\,{\sc i} $\lambda$4471\AA\ to Mg\,{\sc ii} $\lambda$4481\AA\
and derived both
\vs\ and  the inclination angle for a sample of 19 rapidly rotating B stars.
The He lines selected for the present study, however, do not
allow a similar analysis once they all present similar ionization potentials
and thus probably formed  over a narrow range of latitutes.

The values of \vs\ achieved with our calibration, as well as a comparison among
both metods and also with previous results from the literature (whenever possible)
are in the next section.

\section{\vs\ Results}

The FWHM of the three studied He\,{\sc i} lines were measured
in the same way as the theoretical He\,{\sc i} profiles in all target star spectra. 
The stellar \vs\ values obtained using the calibration discussed in this study
are presented in Table~\ref{main}: we list the sample stars with their respective clusters 
or associations in the first two columns and their estimated effective temperatures in column 3 
(from Daflon \& Cunha 2004 and references therein). The \vs's derived from the synthesis of 
metal lines are listed in column 4. Columns 5-10 give the FWHM's measured from the observed 
He\,{\sc i} lines plus the respective \vs\ derived using the He\,{\sc i} calibration; 
and in column 11 we list the average \vs\ for each star  $<V \sin i_{He}>$ with the dispersions, 
which were computed from the individual He\,{\sc i} lines measurements. 

\begin{figure}
\epsscale{.80}
\plotone{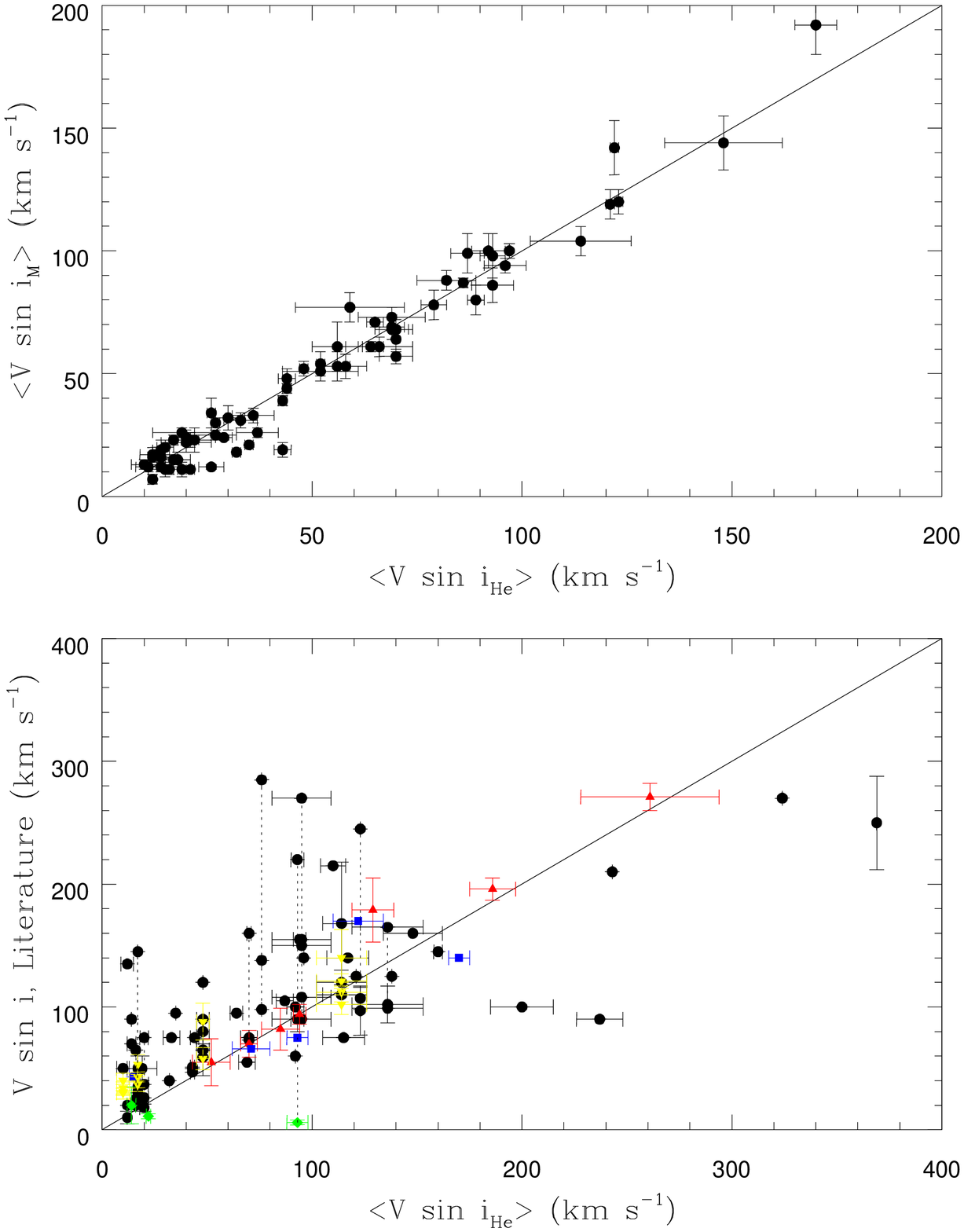}
\caption{{\it Top}: Comparison between \vs\ derived from metal and He\,{\sc i} lines.
{\it Bottom}: Comparison between mean \vs\ from He\,{\sc i} lines and values from the literature.
Black crosses are for \citet{ges00}; green diamonds, \citet{gum98}; red triangles, \citet{heg06a}; 
blue squares, \citet{duf06}; and yellow upsidedown triangles, \citet{sdh07}. \label{comp}}
\end{figure}

A comparison of the \vs's obtained with the two methods discussed in this paper (from synthesis 
of weak metal lines and using the calibration of synthetic FWHM of He\,{\sc i} lines) is shown 
for the sharp lined star sample in the top panel of Figure~\ref{comp}. The \vs\ values are in 
excellent agreement with no systematic differences between the two sets of results: the mean 
difference between $<v \sin i_M>$ and $<v \sin i_{He}>$ is $ -0.4\pm7.7$ \kms. 
Considering each He\,{\sc i} line separetely, the worst agreement with  $<v \sin i_M>$
is found for the He\,{\sc i} line at 4026\AA, which deviates from $<V \sin i_M>$ by roughly +3 \kms. 
This result might perhaps suggest that is preferable to use
only the He\,{\sc i} lines at 4388\AA \, and at 4471\AA \, in \vs\ estimates.
Another recent study also evaluated the differences between \vs's obtained from
the metal lines and He\,{\sc i} profiles. \citet{sdh07} 
used the Fourier method and FWHM of metal and helium lines in order to determine
\vs\ and found that the widths of He lines tend to produce higher 
\vs\ values than the metal lines by a factor of 10-20\%. Our methods for deriving \vs,
however, yield good agreement between the different indicators.
 
A comparison with other \vs\ results from the literature is shown in the bottom panel of 
Figure~\ref{comp}.  The different studies from the literature use different methods to 
obtain \vs\ and these probably have different levels of accuracy.
The \vs\ values from the Fourier method and FWHM by Sim\'on-D\'iaz \& Herrero
(2007, yellow upsidedown triangles) are on average slightly higher than ours, with a mean 
difference of the order of 18 \kms. 
Gummersbach et al. (1998; green diamonds) did an abundance analysis for a sample
of sharp lined stars and we have 3 stars in common with that study: the comparison with our results is
good for two stars, but for BD$-$13\arcdeg4921, we found $<v \sin i_M>= 86 \pm 7$ \kms\ 
and $<V\sin i_{He}>=93\pm5$\kms\ and they find \vs= 6 \kms. 
Dufton et al. (2006, blue squares) also analyzed this star and found \vs= 75 \kms.
Other stars in Dufton et al.'s study are also in good agreement with our results.
A comparison with the results from Huang \& Gies (2006; red triangles), who analyzed the same 
He\,{\sc i} lines as here,  also indicates good agreement: $<$This study $-$ Huang \& Gies $>$ = 
10.0 $\pm$ 18.3 \kms.
Many stars in our sample have \vs\ values in the recent compilation by Glebocki \& Stawikowski 
(2000); these stars are represented as black crosses in the bottom panel of Figure 5.
Since some sample stars have multiple entries in the Glebocki \& Stawikowski catalogue, 
the different \vs\ estimates in each case are connected by the dashed vertical lines in the 
figure. It is clear that the different values in the catalogue are sometimes quite discrepant: 
for example, for HD101436 the catalogue lists \vs=98 , 138 and 235 \kms\, and we find a \vs\ 
value of 76 \kms.  In general, the \vs\ results from the studies based on profile fitting 
agree reasonably well with our \vs\ determinations while the most discrepant points in the 
figure are from the the compilation of Glebocki \& Stawikowski, including results obtained 
with different methods and from spectra of different resolutions, which probably lead
to the large scatter in the multiple results for a given star.

The uncertainties in the \vs\ values derived from the He\,{\sc i} lines 
can be estimated by considering the errors in the measurements of FWHM and
from evaluating the effects of variations in the adopted effective temperature.
In order to estimate the uncertainties in the \vs's, we varied each parameter 
while keeping the other constant.  For a given $T_{\rm eff}$=25,000K and a FWHM=6 \AA, 
the changes in  \vs\  resulting from variations of 10\% in the measured full-width at half 
maximum are similar for the three studied He\,{\sc i} lines:  about 10-12\%.
Considering  FWHM=4 \AA\ and changing $T_{\rm eff}$ from 20,000 to 25,000K, 
we found that the obtained \vs's are increased by 11\%, 3\% and 8\%, for the 3 He\,{\sc i} lines, 
respectively.

The helium profiles are sensitive to surface gravity, being broader for larger values of $\log g$.
Since our calibration was derived for a fixed $\log g$=4.0, we studied the effects of different 
$\log g$ values for a given $T_{\rm eff}$=25,000K. We computed profiles for a complementary grid 
with  $\log g$=3.5 and 4.5 and measured their FWHM consistently. 
Considering a FWHM=$\sim$3\AA, the differences among \vs\ values obtained from profiles with 
$\log g=4.0\pm0.3$ dex are negligible and reach 10--15\%  for $\delta \log g=0.5$ dex.
This result suggests that our calibration can be safely used for main sequence stars with 
$\log g$=3.7--4.3. We note, however, that for the sharpest He lines the effect of gravity 
on the determination of \vs\ may be larger, especially for the line 4026\AA: for this line, 
in the worst case, the derived \vs's can differ $\sim$50\% for  $\delta \log g=0.5$ dex. 
Thus, the use of our calibration results for stars with very sharp
lines and gravities outside the range 3.7 $< \log g <$ 4.3 is not
recomended.   
 
\section{Discussion}

\subsection{The Sample \vs\ Distribution}

\begin{figure}
\epsscale{.80}
\plotone{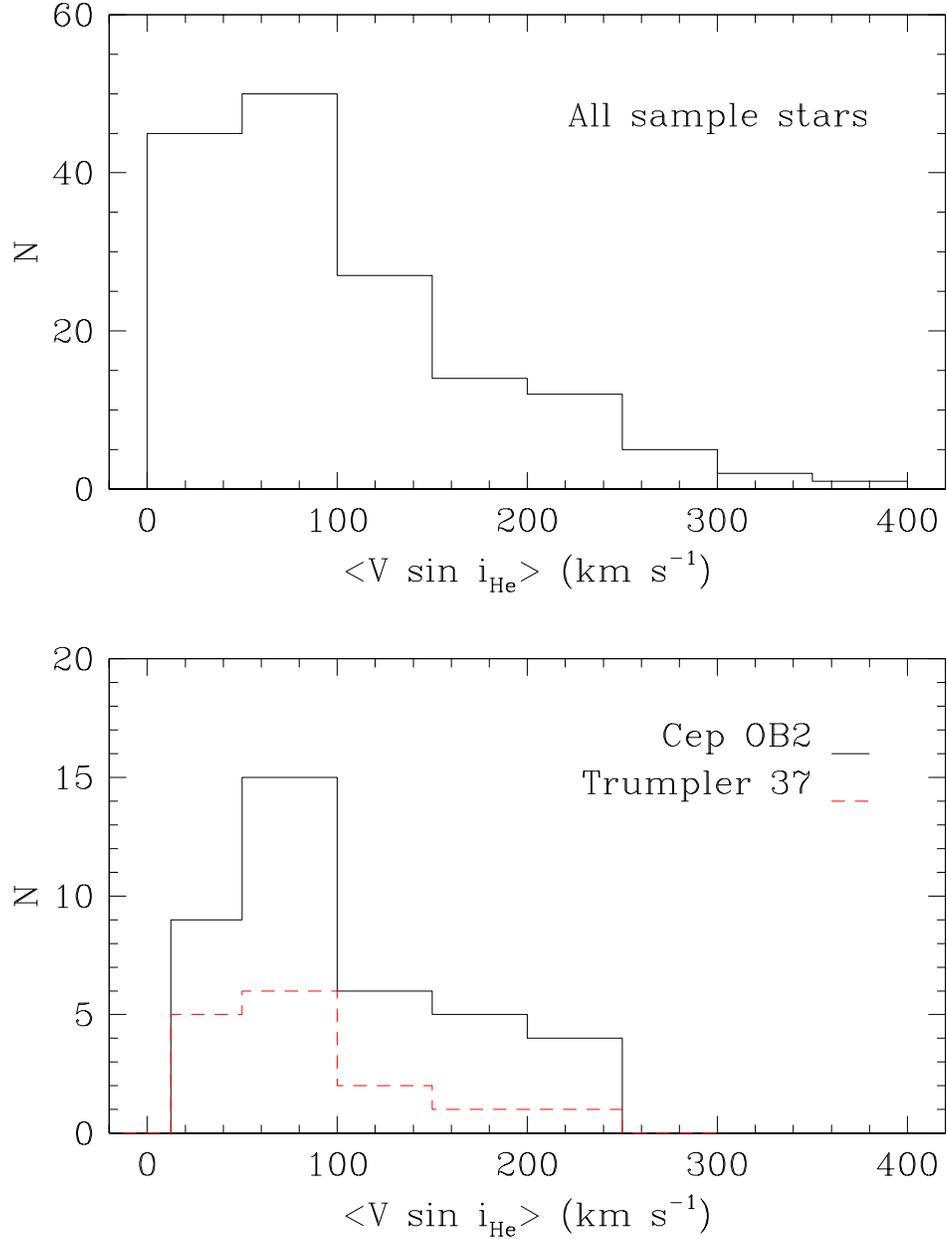}
\caption{{\it Top:} The distributions of \vs\ values for our sample. The
\vs's were obtained with a calibration of synthetic FWHM of He\,{\sc i} lines 
versus stellar projected rotational velocities. 
{\it Bottom: } Distribution of \vs\ for the 40 stars in Cep OB2 association (black histogram) 
and the open cluster Tr 37 (red histogram). \label{hist}}
\end{figure}

The histogram with the \vs\ distribution for the sample of OB stars studied in this paper
(belonging to a variety of clusters and associations) is shown  in the top panel of Figure~\ref{hist}.
The \vs's shown in the histogram were all obtained from the He\,{\sc i} lines.
It is important to note, however, that the main purpose for the observed sample analyzed here
was to construct an abundance database for the study of metallicity gradients in the Milky Way disk.
In this context, \citet{pap7} tried to maximize the number of targets with sharp lines, as 
much as possible, by conducting searches in \vs\ catalogues for OB stars with low values of 
rotational projected velocities before the observing runs. Our observed sample is therefore 
not representative of a randomly selected sample of OB stars in clusters and associations 
and is biased by having a larger number of stars with small projected rotational velocities.
Even so, the average \vs\ for our whole sample of 101$\pm$76 \kms is only 
slightly lower (but still consistent within the standard deviations) than the mean \vs\ 
obtained by \citet{alg02} for   a sample of main sequence B0-B2 stars: 127$\pm$8 \kms.  
This agreement is probably the result of the fact that the the latter sample included mainly 
field stars and members of unbound associations, two groups that also contain
large numbers of slowly rotating stars \citep{wol07}.

\subsection{The Cep OB2 Association}

The Cep OB2 association merits a separate discussion in this study because the observations
conducted for this association are fairly complete down to magnitude V=10
and are without any bias in the selection of targets. We observed
40 targets between spectral types O9-B3 representing all the main-sequence stars (within
these spectral types) listed in \citet{ges92}. Two stars have been discarded from our 
sample due to the very high $T_{\rm eff}$ ($> \sim$35,000K) obtained. 
The lines for two double-lined spectroscopic binaries (HD~208905 and HD~239738) could be resolved, 
resulting in \vs\ determinations for the two components in each system, which were considered
separately.  Three B stars out of our original list are not included in Table~\ref{main} because
a defect in the detector prevented measurement of He\,{\sc i} 4471 \AA\ and the measurements of 
the He\,{\sc i} 4388 \AA\ are rather uncertain, due to bad quality data. (The He line at 4026\AA\
is not covered in the spectra of the northern sample).  In order to have a
complete sample for comparison with data in the literature on other clusters
and associations we have estimated approximate \vs\ values from
He\,{\sc i} 4388 \AA\ as follows:  HD~204827, 102 \kms; HD~206081, 245 \kms; HD~239869, 229 \kms.
Our final sample thus comprises 40 stars: 16 stars belong to the youngest cluster
Trumpler 37, the nucleus of the OB association (with an age of 11 Myr, Dias et al. 2002) 
and 24 stars are from a larger area which includes the
more sparse cluster NGC 7160 (with an age of 19 Myr, Dias et al. 2002).

The histograms in the bottom panel of Figure~\ref{hist} show the \vs\ distributions of
the stars in the Cep OB2 association (represented by the black histogram) and in Tr 37
(represented by the red histogram).  The two distributions are not significantly
different, especially given the small number of stars in each sample.

A recent paper has presented evidence that the \vs\ distribution in clusters
depends on the environment in which
the stars formed, with the differences being much larger at low velocities.  The
sense of the difference is that a larger fraction of stars with low rotation
are found in low density regions as compared with stars formed in
high density regions \citep{wol07}.  The proportion of slow rotators
among field stars is still higher.

\begin{figure}
\epsscale{.80}
\plotone{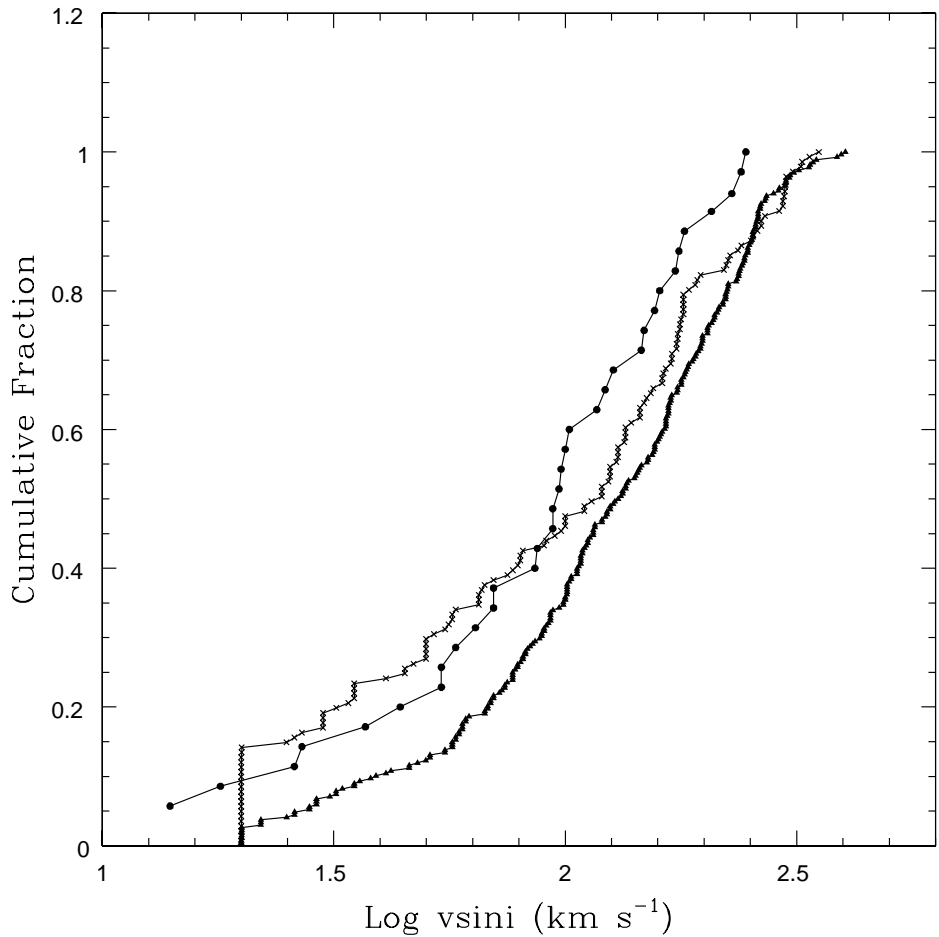}
\caption{Comparison of the cumulative distribution of \vs\ for 35 B stars in
Cep OB2 (filled circles) with the distribution for 141 stars in low density
regions (crosses) and 268 stars in high density regions (triangles).  The
distribution of \vs\ for the stars in Cep OB2 is more similar to the
distribution of stars formed in low density regions but the distinction is
only marginally significant because of the relatively small number of stars in
Cep OB2.\label{cep}}
\end{figure}

If we assume that Cep OB2 occupies the region within {\it l}=96 -- 108$\arcdeg$ and 
{\it b}=$-$1 -- +12$\arcdeg$ with a distance from the Sun of 615 pc \citep{dez99},
the median radius of the association is 64 pc. The total mass of
stars earlier than B3 is $4\times 10^3 M_\sun$ \citep{sim68} and
the stellar density in Cep OB2 is thus 0.02 $M_\sun pc^{-3}$. According to these parameters
and following  \citep{wol07}, Cep OB2 can be classified as  an {\it unbound} stellar ensemble,
as already suggested by \citet{kun86}.  Figure~\ref{cep} compares the cumulative
distribution of \vs\ for all of the stars in Cep OB2, Trumpler 37, and NGC
7160 combined with the distributions for stars in low density, unbound associations and high density, 
bound clusters taken from \citet{wol07}.  For this comparison, we have eliminated the five
stars in Table~\ref{main} that are hotter than the temperature range included in the Wolff et al. survey.
As the figure shows, the distribution for Cep OB2 is similar to that of the stars in low density regions.  
A K-S test indicates that the probability that the distributions for Cep OB2 and the open associations 
are drawn from the same parent
 sample is 0.60.  For both groups, more than 20\% of the stars have projected rotational 
velocities that are less than 50 \kms\ and about 50\% have apparent rotational velocities less
than 100 \kms. However, a K-S test shows that there is also a 9\% probability that the distribution 
for Cep OB2 is drawn from the same parent sample as the bound clusters in Fig.~\ref{cep}. 
Given the large intrinsic spread in the distribution of \vs\ in any group of B stars, a sample even 
as large as we have observed in Cep OB2
is not enough to make a definitive test of a potential relationship between
environment and rotation.  (The probability that the distributions for the stars in low and high density 
regions shown in Fig.~\ref{cep} are drawn from the same parent
population is only 0.001.  The difference between these two groups is more significant than the difference 
between Cep OB2 and the high density group because of the much larger sample size of the stars in low density 
regions--141 stars in low density regions and only 35 stars in Cep OB2.)

\subsection{Summary and Conclusions}

We present \vs\ values for a sample of 156 main sequence OB stars in the Galactic disk.
The \vs's were derived from a calibration for the FWHM of theoretical profiles of He\,{\sc i}
lines at  $\lambda\lambda$4026, 4388, and 4471 \AA\ as a function of the stellar rotation. 
 For a subsample of 69 sharp-lined stars, we also derived
\vs\ from non-LTE synthesis of metal lines. The \vs\ values obtained 
from metal and helium lines show good agreement, and the values we derive
are in good agreement with published rotational velocities. Thus, we conclude 
that our calibration is an useful and rapid tool for estimating the projected rotational velocities
for O9 to B5 main-sequence stars that are rotating at \vs\ up to about 300\kms. 
For most of the stars in this study, our distribution of \vs\
is biased toward low \vs\ and is not representative for main sequence B stars 
sampled randomly. This effect is related to the selection criteria originally
imposed in order to study abundance gradients in the Galactic 
disk. The one exception is the association, Cep OB2, where all the main sequence stars earlier
than B3 were observed. The distribution of \vs\ for this unbound association
is consistent with the distribution for stars in other unbound associations
studied by \citet{wol07}, thereby adding support for the hypothesis that
the distribution of rotational velocities for early B stars depends on the
environment in which they formed.
 
\acknowledgments

\clearpage

\clearpage

\begin{deluxetable}{llccccccccc}
\tabletypesize{\footnotesize}
\rotate
\tablecolumns{12}
\tablewidth{0pc}
\tablecaption{$V \sin i$ Results\label{main}}
\tablehead{
\colhead{Cluster} & \colhead{Star} & \colhead{$T_{\rm eff}$} & \colhead{$<V \sin i_M>$}   & \colhead{W$_{4026}$} &
\colhead{$V\sin i_{4026}$} & \colhead{W$_{4387}$} & \colhead{$V\sin i_{4387}$}  & \colhead{W$_{4471}$} & \colhead{$V\sin i_{4471}$} 
   & \colhead{$<V\sin i_{He}>$}  \\
\colhead{ } & \colhead{ }    & \colhead{ (K) } & \colhead{ (km s$^{-1}$)}   & \colhead{ } &
\colhead{(km s$^{-1}$)}    & \colhead{ }   & \colhead{(km s$^{-1}$)}    & \colhead{ } & \colhead{(km s$^{-1}$)} 
   & \colhead{(km s$^{-1}$)}  }
\startdata
Ara OB1  & HD 149065      & 21540 & 24$\pm$1  & \nodata & \nodata & 1.5     & 31      & 1.5     & 28      & 29$\pm$2   \\
         & HD 153106      & 25510 & \nodata   & 4.5     & 180     & 4.0     & 163     & 4.5:    & 161     & 168$\pm$10 \\
Cen OB1  & HD 100368\tablenotemark{\dagger} & 22000 & \nodata & \nodata & \nodata & 1.2: & 14   & 1.2:    & 9       & 11$\pm$4    \\
         & HD 101794      & 23970 & \nodata   & 5.4     & 227     & 5.7     & 231     & 5.5     & 207     & 224$\pm$15 \\
         & HD 102415      & 30290\tablenotemark{1} & \nodata & 7.0:  & 318  & \nodata & \nodata & 7.6     & 317   & 318$\pm$1  \\
         & HD 102463\tablenotemark{\dagger} & 20000 & \nodata & 2.8  & 65   & 2.3   & 72      & 2.7     & 78      & 72$\pm$6    \\
Cep OB2  & HD 203374      & 30000 & \nodata   & \nodata & \nodata & 5.5: & 234 & \nodata & \nodata & 234 \\
         & HD 204150      & 26450 & \nodata   & \nodata & \nodata & 3.6     & 145     & 4.2     & 147     & 146$\pm$2 \\
         & HD 205794      & 26890 & 12$\pm$2  & \nodata & \nodata & 1.4 & 16   & 1.2  & 12    & 14$\pm$3 \\
         & HD 205948      & 24350 & 144$\pm$11& \nodata & \nodata & 3.9     & 157     & 4.1     & 138     & 148$\pm$14 \\
         & HD 206183      & 33310\tablenotemark{1} & 16$\pm$2 & \nodata & \nodata & 1.0  & 12  & \nodata & \nodata & 12 \\
         & HD 206267A     & 33330\tablenotemark{1} & \nodata & \nodata & \nodata & 2.7: & 105  & 2.6:    & 85  & 95$\pm$14 \\
         & HD 206267C     & 26760 & \nodata & \nodata   & \nodata & 4.2     & 174      & 4.8    & 178   & 176$\pm$3 \\
         & HD 206267D     & 26100 & 44$\pm$2 & \nodata  & \nodata  & 1.6    & 43    & 1.7     & 45    & 44$\pm$1 \\
         & HD 206327      & 21900 & 15$\pm$3 & \nodata  & \nodata  & 1.3     & 20    & 1.3     & 15    & 18$\pm$3 \\
         & HD 207017\tablenotemark{\dagger} & 19410 & \nodata & \nodata & \nodata  & 4.8     & 194   & 4.8    & 167   & 181$\pm$19 \\
         & HD 207308      & 23250 & \nodata& \nodata   & \nodata   & 4.8:    & 198  & 5.7     & 216   & 207$\pm$13 \\
         & HD 207538      & 32190\tablenotemark{1} & 39$\pm$2 & \nodata  & \nodata  & 1.5  & 42   & 1.6   & 44   & 43$\pm$1  \\
         & HD 207951      & 20650 & 87$\pm$2 & \nodata  & \nodata  & 2.5 & 86  & \nodata & \nodata & 86 \\
         & HD 208266      & 24840 & \nodata & \nodata   & \nodata  & 3.2:    & 124   & 3.5:    & 110   & 117$\pm$10 \\
         & HD 208440      & 26470 & \nodata & \nodata   & \nodata  & 2.5     & 92    & 3.1     & 96   & 94$\pm$3  \\
         & HD 208905a\tablenotemark{\ddagger} & 27460 & \nodata & \nodata & \nodata  & 1.4   & 33   & 1.6     & 41    & 37$\pm$5  \\
         & HD 208905b\tablenotemark{\ddagger} & 27460 & \nodata & \nodata & \nodata  & 1.7   & 54    & 2.0:    & 62    & 58$\pm$6  \\
         & HD 209339      & 31250\tablenotemark{1} & 98$\pm$9 & \nodata  & \nodata  & 2.5 & 96  & 2.8  & 91  & 93$\pm$3 \\
         & HD 209454\tablenotemark{\ast}  & 25460 & \nodata & \nodata & \nodata  & 4.2:    & 173  & \nodata & \nodata   & 173  \\
         & HD 213023      & 26050 & \nodata & \nodata   & \nodata  & 2.5      & 91   & 3.1:   & 96   & 94$\pm$3 \\
         & HD 213757      & 20000 & \nodata & \nodata   & \nodata  & 4.1     & 162  & 4.7     & 159   & 160$\pm$2 \\
         & HD 235618      & 27180 & 100$\pm$3 & \nodata & \nodata  & 2.6     & 97   & 3.1     & 97   & 97$\pm$0 \\
         & HD 239649      & 20000 & \nodata & \nodata   & \nodata  & 3.8     & 148  & 4.8     & 164   & 156$\pm$11 \\
         & HD 239676\tablenotemark{\ast}  & 24720 & \nodata & \nodata & \nodata  & 2.6:    & 95   & 3.3:    & 102   & 98$\pm$5 \\
         & HD 239681      & 26830 & 142$\pm$11& \nodata & \nodata  & 3.1    & 121    & 3.7     & 123   & 122$\pm$1  \\
         & HD 239710      & 21900 & 64$\pm$5 & \nodata  & \nodata  & 2.2    & 70   & 2.4    & 70   & 70$\pm$0 \\
         & HD 239712      & 18870 & \nodata & \nodata   & \nodata  & 5.9:  & 243  & \nodata & \nodata   & 243 \\
         & HD 239724      & 24790 & 34$\pm$6 & \nodata   & \nodata & 1.3   & 26   & \nodata & \nodata   & 26  \\
         & HD 239725      & 20540 & \nodata & \nodata   & \nodata  & 3.3   & 125  & 4.1     & 130   & 127$\pm$4  \\
         & HD 239729      & 28450 & 99$\pm$8 & \nodata   & \nodata & 2.3    & 84   & 2.8     & 89     & 87$\pm$4  \\
         & HD 239738a\tablenotemark{\ddagger} & 21060 & \nodata & \nodata   & \nodata  & \nodata & \nodata & 2.0: & 54 &  54  \\
         & HD 239738b\tablenotemark{\ddagger} & 21060 & \nodata & \nodata   & \nodata  & 2.0 & 58 & 1.9 & 50 & 54$\pm$6  \\
         & HD 239742      & 22470 & 30$\pm$2 & \nodata   & \nodata   & 1.4   & 27   & \nodata   & \nodata   & 27  \\
         & HD 239743      & 21580 & 17$\pm$3 & \nodata   & \nodata   & 1.2    & 14  & 1.2       & 10   & 12$\pm$3  \\
         & HD 239745      & 27340 & 61$\pm$2 & \nodata   & \nodata   & 1.9     & 62  & 2.1       & 66  & 64$\pm$3  \\
         & HD 239748      & 27480 & 68$\pm$4 & \nodata   & \nodata   & 2.0   & 67   & 2.3       & 73 & 70$\pm$4 \\
         & HD 239767\tablenotemark{\ast}  & 29940 & \nodata & \nodata   & \nodata   & 2.6:  & 100  & \nodata & \nodata & 100 \\
Cep OB3  & BD +62\arcdeg2125& 23480 & 48$\pm$4 & \nodata  & \nodata   & 1.7 & 45  & 1.7  & 42    & 44$\pm$2 \\
         & BD +62\arcdeg2127& 22630 & \nodata & \nodata  & \nodata   & 2.6 & 92 & \nodata   & \nodata  & 92 \\
         & HD 217657      & 27950 & 13$\pm$2 & \nodata   & \nodata   & 1.1 & 13   & 1.1      & 8   & 10$\pm$3  \\
         & HD 218342      & 30020\tablenotemark{1} & 31$\pm$3 & \nodata   & \nodata & 1.3 & 30  & 1.5   & 37   & 33$\pm$4  \\
Cyg OB3  & BD +35\arcdeg3956& 24840 & \nodata & \nodata  & \nodata   & 6.5  & 273  & \nodata & \nodata & 273 \\
         & HD 191566      & 27290 & \nodata & \nodata   & \nodata   & 2.6 & 96 & \nodata  & \nodata  & 96  \\
         & HD 191567\tablenotemark{\ast} & 22520 & \nodata& \nodata& \nodata& 1.6 & 20 & \nodata& \nodata& 20  \\
         & HD 227460      & 27060 & 11$\pm$3 & \nodata   & \nodata   & 1.5 & 19 & \nodata  & \nodata  & 19  \\
         & HD 227586      & 27830 & 21$\pm$1 & \nodata   & \nodata   & 1.6 & 35 & \nodata  & \nodata  & 35  \\
         & HD 227621      & 23230 & \nodata & \nodata   & \nodata   & 2.8   & 103    & \nodata   & \nodata   & 103  \\
         & HD 227696      & 29100 & 120$\pm$5 & \nodata   & \nodata   & 3.1 & 123   & \nodata   & \nodata   & 123  \\
         & HD 227757 & 32480\tablenotemark{1} & 26$\pm$2 & \nodata & \nodata & 1.3 & 15 & 1.3 & 24 & 19$\pm$7\\
         & HD 227877      & 23260 & \nodata & \nodata   & \nodata & 3.7 & 147 & \nodata & \nodata & 147 \\
         & HD 228199      & 29870 & 104$\pm$6 & \nodata   & \nodata   & 2.7   & 105     & 3.0  & 122   & 114$\pm$12 \\
Cyg OB7 & BD +44\arcdeg3594& 30620\tablenotemark{1}& \nodata& \nodata& \nodata& 4.6& 195& \nodata& \nodata&195\\
         & HD 197512      & 23570 & 12$\pm$2 & \nodata  & \nodata   & 1.5 & 14 & \nodata   & \nodata   & 14 \\
         & HD 198931      & 23410 & \nodata & \nodata   & \nodata   & 8.7: & 369 & \nodata   & \nodata   & 369 \\
         & HD 199579      & 32930\tablenotemark{1} & \nodata & \nodata   & \nodata & 2.0  & 70  & 2.1   & 69  & 70$\pm$1 \\
         & HD 201666      & 19900 & \nodata & \nodata   & \nodata   & 3.6 & 138   & \nodata   & \nodata   & 138 \\
         & HD 202163      & 18560 & \nodata & \nodata   & \nodata   & 4.7 & 189 & \nodata & \nodata & 189 \\
         & HD 202253      & 22750 & 52$\pm$3 & \nodata   & \nodata   & 1.8:   & 48      & \nodata   & \nodata   & 48 \\
         & HD 202347      & 23280 & 119$\pm$6 & \nodata   & \nodata   & 3.2    & 121   & \nodata   & \nodata   & 121 \\
IC 2944  & HD 101070\tablenotemark{\dagger} & 29550 & \nodata & 2.7     & 87      & 2.1    & 75    & 2.9:      & 92   & 85$\pm$9  \\
         & HD 101084      & 26750 & \nodata & 3.7    & 136    & \nodata & \nodata& 3.7     & 123  & 129$\pm$10  \\
         & HD 101223\tablenotemark{\dagger} & 30800\tablenotemark{1} & \nodata & 2.4:  & 71  & 1.9  & 65   & 2.2  & 73 & 70$\pm$4  \\
         & HD 101413      & 28430 & \nodata & 3.4    & 124    & 3.6    & 148    & \nodata & \nodata& 136$\pm$17  \\
         & HD 101436\tablenotemark{\dagger} & 33270\tablenotemark{1} & \nodata& \nodata& \nodata& \nodata& \nodata & 2.3: & 76 & 76 \\
         & HD 308810      & 26400 & 51$\pm$4 & 2.1  & 43    & 1.8   & 54   & 2.0    & 60    & 52$\pm$9  \\
         & HD 308813      & 33320\tablenotemark{1} & \nodata & 4.4   & 183   & 4.2   & 177 & 5.1  & 198  & 186$\pm$11  \\
         & HD 308817      & 22940 & 57$\pm$3 & 2.7  & 68    & 2.1   & 66     & 2.5   & 74  & 70$\pm$4  \\
         & HD 308833      & 23340 & \nodata & 6.5   & 284    & 5.7:    & 237   & \nodata   & \nodata & 261$\pm$33  \\
Lac OB1  & HD 214167      & 26720 & 18$\pm$1 & \nodata   & \nodata   & 1.4  & 32  & \nodata   & \nodata   & 32  \\
         & HD 214680      & 33690\tablenotemark{1} & 23$\pm$2 & \nodata   & \nodata   & 1.1  & 17  & \nodata & \nodata & 17  \\
         & HD 216916      & 23520 & 11$\pm$2 & \nodata   & \nodata   & 1.2   & 16   & \nodata   & \nodata   & 16  \\
         & HD 217227      & 19000 & 15$\pm$2 & \nodata   & \nodata   & 1.3   & 17   & \nodata   & \nodata   & 17  \\
         & HD 217811      & 19070 & 7$\pm$2 & \nodata   & \nodata   & 1.2   & 12    & \nodata   & \nodata   & 12  \\
         & HD 218674      & 18840 & \nodata & \nodata   & \nodata   & 7.8   & 324   & \nodata   & \nodata   & 324  \\
Mon OB2  & HD 46106       & 29150 & \nodata & 3.2:    & 114    & 2.7     & 104     & 3.6:     & 122  & 114$\pm$9  \\
         & HD 46202       & 31500\tablenotemark{1} & 24$\pm$3 & \nodata & \nodata & 1.1 & 19 & 1.3 & 22  & 20$\pm$2 \\
         & HD 46966       & 30230\tablenotemark{1} & \nodata & 2.0   & 44    & 1.6    & 48   & 1.7 & 51  & 48$\pm$3 \\
         & HD 47360       & 26250 & \nodata & 5.3    & 226     & 5.6    & 237     & 6.3   & 248   & 237$\pm$11 \\
         & HD 47417       & 31550\tablenotemark{1} & \nodata & 3.4 & 128   & 3.2  & 129    & 3.7   & 129  & 129$\pm$1  \\
         & HD 259105      & 27750 & \nodata & 4.6   & 190     & 4.5      & 188   & 5.0     & 190  & 189$\pm$1  \\
NGC 1893 & S2R2N43        & 24020 & 54$\pm$5 & \nodata   & \nodata   & 1.8   & 52   & \nodata   & \nodata   & 52  \\
         & S2R3N09        & 26160 & 11$\pm$3 & \nodata   & \nodata   & 1.2  & 18  & 1.2  & 13  & 15$\pm$4  \\
NGC 2414 & LS 400         & 30000 & \nodata & 5.0     & 216    & 5.3   & 226      & 5.4    & 213  & 218$\pm$7  \\
         & LS 404         & 23260 & 32$\pm$5 & 2.0   & 26      & 1.5    & 33    & 1.5     & 25  & 30$\pm$4  \\
         & LS 427a\tablenotemark{\ddagger}        & 28790 & \nodata & 3.4 & 124 & 3.2 & 128  & 3.6: & 121 & 125$\pm$3  \\
         & LS 427b\tablenotemark{\ddagger}        & 28790 & \nodata & 3.8 & 147 & 3.6 & 147  & 4.0: & 142 & 145$\pm$3  \\
         & LS 428         & 28140 & 33$\pm$3 & 1.9 & 32 & 1.4 & 34 & 1.6  & 42  & 36$\pm$5   \\
         & LS 531         & 26070 & \nodata & 4.5  & 181   & 4.5  & 181    & 4.8       & 177  & 181$\pm$5 \\
NGC 2439 & CD -32\arcdeg4257a\tablenotemark{\ddagger}& 28560 & \nodata  & \nodata & \nodata  & 5.0 & 212 & 5.5 & 215  & 213$\pm$3  \\
         & CD -32\arcdeg4257b\tablenotemark{\ddagger}& 28560 & \nodata  & \nodata  & \nodata & 6.2 & 264 & 6.5 & 263  & 263$\pm$1  \\
         & CPD -33\arcdeg1682\tablenotemark{\ast}& 33210\tablenotemark{1} & \nodata& 2.3 & 64 & 2.0: & 70 & 2.0 & 65 & 67$\pm$3 \\
         & HD 61851       & 28060 & \nodata & 4.0:   & 157    & 3.7:   & 151    & 4.1     & 145   & 151$\pm$5  \\
         & LS 709         & 31410\tablenotemark{1} & \nodata & 4.5  & 191    & 4.8   & 204   & 5.8   & 231   & 209$\pm$20  \\
NGC 3576 & HD 97499       & 28590 & \nodata & 4.9   & 208   & 4.8      & 203     & 5.0:      & 191   & 201$\pm$8  \\
NGC 4755 & CPD -59\arcdeg4532& 23610 & 71$\pm$2 & 2.6  & 64     & 2.1   & 67    & 2.2     & 65   & 65$\pm$2  \\
         & CPD -59\arcdeg4535& 22930 & 68$\pm$4 & 2.7   & 68    & 2.1   & 66    & 2.5    & 74  & 69$\pm$4  \\
         & CPD -59\arcdeg4544& 24330 & 69$\pm$5 & 2.6   & 66    & 2.1  & 68    & 2.4    & 72  & 69$\pm$3  \\
         & CPD -59\arcdeg4557& 23240 &\nodata& 3.7 & 127     & 2.9     & 108   & 4.0      & 131  & 122$\pm$12  \\
         & CPD -59\arcdeg4560& 22390 & 192$\pm$12 & 4.5 & 172    & 4.4   & 174   & 4.7   & 164   & 170$\pm$5  \\
NGC 6031 & NGC 6031 12\tablenotemark{\ast} & 19050 & \nodata & 1.9   & 17  & \nodata  & \nodata    & 1.4    & 24  & 21$\pm$5  \\
         & NGC 6031 40    & 19010: & \nodata & 3.5   & 108     & \nodata  & \nodata    & 3.3   & 98  & 103$\pm$6  \\
         & NGC 6031 74    & 21600: & \nodata & 2.2:  & 34     & \nodata  & \nodata    & 1.6   & 34   & 34$\pm$0  \\
NGC 6204 & CPD -46\arcdeg8206& 15000: & \nodata & 1.8:  & 39   & 1.3:     & 22    & 1.1:      & 26    & 29$\pm$9  \\
         & HD 150627       & 29620 & \nodata & 3.8    & 149   & 3.6      & 148     & 4.0       & 143   & 147$\pm$3  \\
         & LS 3719         & 24120 & 77$\pm$6 & 2.3  & 50   & 1.8   & 52     & 2.5    & 74  & 59$\pm$13  \\
NGC 6604 & BD -12\arcdeg4978& 27750& 61$\pm$4 & 2.3  & 58   & 2.0   & 68     & 2.3   & 73  & 66$\pm$8  \\
NGC 6611 & BD -12\arcdeg5074& 26210& 19$\pm$3 & 2.1  & 41     & 1.6    & 43    & 1.7    & 45  & 43$\pm$2  \\
         & BD -13\arcdeg4921& 29540& 86$\pm$7 & \nodata   & \nodata   & 2.4   & 90    & 3.0    & 97    & 93$\pm$5  \\
         & BD -13\arcdeg4929& 30600\tablenotemark{1}& \nodata & 2.6:  & 80  & 2.0  & 70   & 1.9   & 62   & 71$\pm$9  \\
         & BD -13\arcdeg4930& 30830\tablenotemark{1} & 20$\pm$3 & \nodata & \nodata & 1.1  & 17  & 1.2  & 13  & 15$\pm$2  \\
         & BD -13\arcdeg4934& 30970\tablenotemark{1} & 80$\pm$6 & \nodata   & \nodata   & 2.4 & 91 & 2.7 & 88 & 89$\pm$2  \\
         & HD 170452      & 29890  & \nodata & 3.2:   & 116   & \nodata  & \nodata    & 3.4   & 114   & 115$\pm$2  \\
NGC 6231 & CPD -41\arcdeg7723& 24920 & 26$\pm$2 & 2.1   & 37    & 1.6   & 42     & 1.5    & 31  & 37$\pm$5  \\
         & CPD -41\arcdeg7730& 24670 & 12$\pm$1 & 1.9 & 23      & 1.4   & 30     & 1.4   & 24    & 26$\pm$3  \\
         & HD 326332      & 27030  & 23$\pm$3 & 1.7     & 21    & 1.2   & 20     & 1.4   & 27  & 21$\pm$5  \\
         & HD 326364      & 29610  & 19$\pm$4 & 1.6   & 13     & 1.1     & 16       & 1.2    & 13   &  14$\pm$2  \\
Pismis 20& Pismis 20 6    & 27650  & \nodata & \nodata & \nodata  & \nodata  & \nodata  & 2.4    & 76   & 76  \\
R 105    & HD 144900      & 27340  & \nodata & 3.0:      & 103    & 2.9:     & 112    & 3.5:      & 114   & 110$\pm$6  \\
         & HD 144970      & 30760\tablenotemark{1}  & \nodata & 3.2   & 116   & 3.1    & 125    & 3.2   & 105  & 115$\pm$10 \\
R 103    & CPD -50\arcdeg9216 & 20040 & \nodata & \nodata & \nodata   & 5.6   & 230   & \nodata   & \nodata   & 230  \\
Sct OB2  & BD -08\arcdeg4617& 31410\tablenotemark{1} & \nodata & 3.1  & 111   & 2.8    & 110  & 2.9 & 95  & 105$\pm$9 \\
         & BD -08\arcdeg4634& 29360 & \nodata & 2.4   & 69     & 2.2      & 80    & 2.5    & 82  & 77$\pm$7  \\
         & HD 172367      & 28640  & \nodata & 5.3:   & 229   & 5.7    & 242    & 6.2:     & 249   & 240$\pm$10  \\
         & HD 172427      & 26360  & 53$\pm$5 & 2.3  & 52     & 1.9   & 60     & 2.0      & 60  & 58$\pm$5  \\
         & HD 172488      & 26530  & 100$\pm$7 & 2.9  & 89     & 2.6  & 97     & 2.9     & 90  & 92$\pm$4  \\
         & HD 173637      & 31950\tablenotemark{1}  & \nodata & 4.5 & 190  & 5.0 & 211 &  \nodata &  \nodata & 200$\pm$15  \\
Sh 2-16  & LS 4381        & 22500  & \nodata & 3.3:   & 100   & \nodata & \nodata & 2.9   & 86  & 93$\pm$10  \\
Sh 2-29  & HD 166192\tablenotemark{\ast} & 28010  & \nodata & 3.7   & 139    & 3.2: & 127     & 3.5:      & 115   & 127$\pm$12  \\
Sh 2-32  & HD 166033      & 27290  & 73$\pm$4 & 2.4  & 63     & 2.0    & 67      & 2.5    & 78   & 69$\pm$8  \\
         & HD 314031      & 27650  & 94$\pm$3 & 2.9    & 93     & 2.5    & 93    & 3.2    & 102  & 96$\pm$5  \\
Sh 2-47  & Sh 2-47 3      & 29870  & 16$\pm$2 & 1.6    & 13   & 1.1  & 16    & 1.2   & 13  & 14$\pm$2   \\
Sh 2-247 & Sh 2-247 1     & 31560\tablenotemark{1}  & 23$\pm$5 & 1.7   & 22  & 1.2   & 23  & 1.3   & 22  & 22$\pm$1  \\
Sh 2-253 & LS 45          & 22830  & 22$\pm$4 & \nodata   & \nodata   & 1.4  & 25   & 1.3  & 16   & 20$\pm$6  \\
         & LS 51          & 32630\tablenotemark{1}  & \nodata & 6.1 & 273   & 6.6:  & 282  & 7.9:  & 275  & 277$\pm$5  \\
Sh 2-284 & HD 48691       & 27870  & 53$\pm$6 & 2.2   & 52  & 1.8   & 57    & 1.9      & 58  & 56$\pm$3 \\
Sh 2-285 & BD -00\arcdeg1491& 29480 & 12$\pm$2 & \nodata & \nodata  & 1.0  & 9       & 1.2     & 13   & 11$\pm$3 \\
Sh 2-289 & Sh 2-289 2     & 25060 & \nodata & \nodata   & \nodata   & 2.3    & 80     & 2.4    & 74  & 76$\pm$5  \\
         & Sh 2-289 4     & 24560 & \nodata & \nodata   & \nodata   & 2.2   & 74      & 1.2    & 58  & 66$\pm$11  \\
Sh 2-301 & LS 212         & 21980 & \nodata & 5.0   & 200    & 5.9  & 245      & 5.7    & 214  & 219$\pm$23 \\
Stock 16 & CPD -61\arcdeg3576& 29100 & \nodata & 2.6: & 80   & 2.3:   & 85    & 2.9:   & 93  & 86$\pm$7 \\
         & CPD -61\arcdeg3579& 27840  & 78$\pm$6 & 2.6 & 76     & 2.2    & 78    & 2.6    & 83  & 79$\pm$3  \\
         & CPD -61\arcdeg3581& 25810 & \nodata & 5.9:   & 257   & 6.3:     & 266     & 6.9:    & 279  & 267$\pm$11 \\
Tr 27    & LS 4257        & 29460  & 88$\pm$4 & 2.5   & 76     & 2.2     & 80    & 2.8   & 89  & 82$\pm$7 \\
         & LS 4264        & 22360  & \nodata & 3.1   & 88     & 2.9:     & 107     & 3.7:      & 115   & 103$\pm$13  \\
         & LS 4271        & 32190\tablenotemark{1}  & 61$\pm$10 & 2.1  & 51   & 1.7    & 54   & 2.0   & 64  & 56$\pm$6  \\
Vul OB1  & BD +24\arcdeg3880& 30410\tablenotemark{1}  & 11$\pm$1 & \nodata & \nodata & 1.1 & 21 & \nodata & \nodata & 21  \\
         & BD +24\arcdeg3843& 23250  & \nodata & \nodata & \nodata   & 4.1  & 166  & 3.4    & 131   & 148$\pm$25  \\
         & HD 344783      & 31010\tablenotemark{1}  & 25$\pm$2 & 1.7  & 22 & 1.3 & 30 & 1.4  & 30  & 27$\pm$5  \\
         & NGC 6823 Hoag6 & 31550\tablenotemark{1} & \nodata & \nodata & \nodata & 2.2  & 81  & 2.6  & 85  & 83$\pm$3  \\
...      & RL 41          & 20350 & \nodata & \nodata & \nodata  & 2.9  & 104  & 3.3  & 96 & 100$\pm$6  \\
\enddata
\tablenotetext{1}{Assumed as Tef=30000K to calculate \vs}
\tablenotetext{:}{Uncertain measurement of FWHM}
\tablenotetext{\dagger}{Asymmetric line profiles: possible spectroscopic binary?} \tablenotetext{\ddagger}{Single-lined spectroscopic binary (SB1)} \tablenotetext{\ast}{Double-lined spectroscopic binary(SB2)}

\end{deluxetable}


\begin{deluxetable}{c|ccc|ccc}
\tablecolumns{7}
\tablewidth{0pc}
\tablecaption{Full-Widths at Half Maximum of theoretical He profiles (in \AA), for R=50,000 \label{grid}}
\tablehead{ \colhead{V $\sin i$} & \colhead{4026\AA} & \colhead{4388\AA} & \colhead{4471\AA} & \colhead{4026\AA} & \colhead{4388\AA} & \colhead{4471\AA} \\ 
 (km s$^{-1}$) & \colhead{ } & \colhead{ }  & \colhead{ } & \colhead{ } & \colhead{ } & \colhead{ }  }

\startdata
& & $T_{\rm eff}$=15,000 K & & & $T_{\rm eff}$=20,000 K   &  \\
0   & 1.04 & 1.02 & 0.65    & 1.74 & 0.99 & 1.09 \\
50  & 2.01 & 1.90 & 1.59    & 2.54 & 1.89 & 1.93 \\
100 & 3.01 & 2.86 & 3.03    & 3.43 & 2.82 & 3.42 \\
150 & 4.10 & 3.92 & 4.38    & 4.25 & 3.84 & 4.53 \\
200 & 5.11 & 4.94 & 5.49    & 5.08 & 4.92 & 5.49 \\
250 & 6.05 & 6.02 & 6.55    & 5.98 & 6.07 & 6.51 \\
300 & 7.01 & 7.34 & 7.61    & 6.92 & 7.22 & 7.55 \\
350 & 7.99 & 8.51 & 8.66    & 7.90 & 8.36 & 8.60 \\
400 & 8.99 & 9.64 & 9.77    & 8.87 & 9.48 & 9.68 \\
\hline
& & $T_{\rm eff}$=25,000 K & & & $T_{\rm eff}$=30,000 K  &   \\
0   & 1.55 & 0.91 & 1.06    & 1.44 & 0.86 & 1.05 \\
50  & 2.30 & 1.74 & 1.82    & 2.08 & 1.63 & 1.69 \\
100 & 3.16 & 2.70 & 3.25    & 2.91 & 2.59 & 3.07 \\
150 & 4.01 & 3.73 & 4.31    & 3.80 & 3.63 & 4.11 \\
200 & 4.87 & 4.81 & 5.31    & 4.71 & 4.71 & 5.13 \\
250 & 5.79 & 5.95 & 6.33    & 5.65 & 5.86 & 6.18 \\
300 & 6.74 & 7.12 & 7.37    & 6.63 & 7.00 & 7.22 \\
350 & 7.71 & 8.25 & 8.45    & 7.62 & 8.14 & 8.32 \\
400 & 8.71 & 9.36 & 9.54    & 8.65 & 9.27 & 9.42 \\
\enddata 
\end{deluxetable}

\end{document}